\documentclass[twocolumn]{jpsj3}

\usepackage{amsmath}
\usepackage{amsfonts}
\usepackage{amssymb}
\usepackage{bm}
\usepackage{tabularx}
\usepackage{color}
\usepackage{graphicx}
\usepackage{txfonts}

\def\SRO{Sr$_2$RuO$_4$}
\def\P{$P$}
\def\0{$_{\parallel [100]}$}
\def\1{$_{\parallel [110]}$}
\def\c{$_{\parallel [001]}$}
\def\HydroP{$P_\mathrm{hydro}$}
\def\Tc{$T_{\mathrm{c}}$}
\def\RuO6{RuO$_6$}
\def\chiAC{$\chi_\mathrm{AC}$}

\title{Higher-\Tc\ Superconducting Phase in \SRO\ Induced by In-Plane Uniaxial Pressure}

\author{Haruka Taniguchi$^{1, 2}$\thanks{E-mail: tanig@iwate-u.ac.jp}, Keigo Nishimura$^2$, Swee K. Goh$^{3, 4}$, Shingo Yonezawa$^2$, and Yoshiteru Maeno$^2$}
\inst{$^1$ Department of Materials Science and Engineering, Iwate University, Morioka 020-8551, Japan \\
$^2$ Department of Physics, Graduate School of Science, Kyoto University, Kyoto 606-8502, Japan \\
$^3$ Department of Physics, The Chinese University of Hong Kong, Shatin, New Territories, Hong Kong, China \\
$^4$ Cavendish Laboratory, University of Cambridge, J J Thomson Avenue, Cambridge CB3 0HE, UK }

\abst{
We revealed that the superconducting transition temperature \Tc\ of the multi-component superconductor \SRO\ is enhanced to 3.3~K under in-plane uniaxial pressure that reduces the tetragonal crystal symmetry.
This result suggests that new superconducting phases with a one-component order parameter are induced.
We have also clarified the in-plane pressure direction dependence of the emergence of this higher-\Tc\ superconducting phase: 
pressure along the [100] direction is more favorable than pressure along the [110] direction.
This result is probably closely related to the direct shortening of the in-plane Ru-O bond length along the pressure direction and the approach of the $\gamma$ Fermi surface to the van Hove singularity under \P\0.
}


\begin{document}
\maketitle

\section{Introduction}
\SRO, with the transition temperature \Tc\ of 1.5~K, is a leading candidate for chiral-$p$-wave spin-triplet superconductors~\cite{Maeno1994, Rice1995, Ishida1998, Mackenzie1998PRL, Duffy2000, Ishida2001, Maeno2012}.
It is deduced from a number of experiments that the $d$ vector describing its superconducting (SC) state has the form  $\bm{d} = \Delta \hat{z} (k_x \pm ik_y)$ ~\cite{Rice1995, Ishida1998, Luke1998, Duffy2000, Ishida2001, Xia2006},
indicating that the spin and orbital parts satisfy $S_z = 0$ and $L_z = \pm 1$, respectively. 
The degeneracy of the $k_x$ and $k_y$ states is a key for the realization of the chiral orbital state.
This degeneracy is expected to be lifted, and a SC double transition occurs
when the in-plane tetragonal symmetry of the crystal structure is broken by uniaxial pressure (UAP) or a magnetic field~\cite{Agterberg1998, Sigrist2002}.
A phase with a one-component SC order parameter ($k_x$ or $k_y$) is expected in a high-temperature region (\Tc$_-$ $< T <$ \Tc$_+$),
whereas a chiral phase with a two-component complex SC order parameter [$\lambda_x(T) k_x + i \lambda_y(T) k_y$] is realized in the low-temperature region ($0 < T <$ \Tc$_-$).
If some magnetic energy is dissipated at the transition between the one-component and two-component SC phases, 
the imaginary part of the AC magnetic susceptibility, \chiAC, exhibits an anomaly at \Tc$_-$ as well as at \Tc$_+$.
Under an in-plane magnetic field, the possibility of SC double transitions has been reported~\cite{Nishizaki2000, Mao2000PRL, Tanatar2001PRB, Deguchi2002, Yaguchi2002, Tenya2006}.
However, recent thermal measurements~\cite{Yonezawa2013, Yonezawa2014} revealing the first-order SC transition without a clear second anomaly have suggested 
that the previously reported anomalies may have originated from the broadening of the first-order transition or from sample mosaicity.
Thus, the existence of double transitions in the in-plane magnetic field is still unclear.
Therefore, whether SC double transitions occur under UAP is essential for determining the SC order parameter of \SRO.
In other words, if double transitions are observed under in-plane UAP, it would be a crucial proof that the SC order parameter of \SRO\ has multiple components.

Experimentally, \Tc\ of \SRO\ is known to be sensitive to lattice distortion.
For example, \Tc\ is suppressed by hydrostatic pressure \HydroP\ at a rate of $dT_\mathrm{c}/dP_\mathrm{hydro} = -0.2 $ K/GPa~\cite{Shirakawa1997, Forsythe2002}.
On the basis of ultrasonic experiments and the Ehrenfest relation,
\Tc\ is predicted to also be suppressed by the in-plane UAP along the [100] direction \P\0\ at a rate of $(1/T_\mathrm{c})(dT_\mathrm{c}/dP_{[100]}) = - (0.85\pm0.05)$~GPa$^{-1}$
~\cite{Okuda2002PhysicaB, Okuda2002JPSJ}.
In contrast, the enhancement of \Tc\ up to about 3~K is observed in \SRO-Ru eutectic crystals~\cite{Maeno1998},
in which superconductivity with higher \Tc\ probably occurs in the \SRO\ part around \SRO-Ru interfaces
as a consequence of the strong lattice distortion due to lattice mismatch~\cite{Maeno1998, Ando1999, Kittaka2009JPSJ-SCregion}.
A similar enhancement of \Tc\ can be caused by out-of-plane UAP \P\c\ in non eutectic \SRO~\cite{Kittaka2010}.
However, because of the technical difficulties in directly measuring the effects of in-plane UAP and uniaxial strain,
the effect of in-plane UAP on \SRO\ has been under discussion~\cite{Ikeda2001, Yaguchi-2014JPS} and the effect of in-plane uniaxial strain on \SRO\ has been reported only recently~\cite{Hicks2014}.

In this study, we have investigated the AC susceptibility of \SRO\ under in-plane UAP along two directions, [100] or [110],
to obtain hints for determining the SC order parameter of \SRO, whose orbital part is expected to have two components under ambient conditions.

\section{Methods}
\begin{figure}[htb]
\begin{center}
\includegraphics[width=3in]{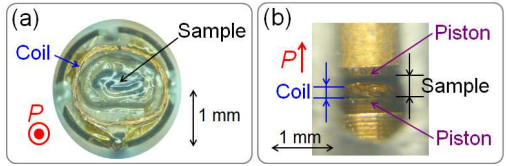}
\end{center}
\caption
{(Color online)
(a) Top view of a pick-up coil and a sample on a piston. Epoxy surrounds the side surface of the sample to prevent the sample from breaking.
Grease (Apiezon, N-type) is spread on the top and bottom surfaces of the sample to improve the thermal link between the sample and the pistons.
(b) Side view of the pick-up coil and the sample through a window on the outer body of a pressure cell.
The coil height is less than the sample height.
}
\label{Coil}
\end{figure}

Single crystals of \SRO\ were grown by the floating-zone method with Ru self-flux~\cite{Mao2000MRB}.
The directions of the crystal axes were determined by the Laue method.
The samples for \P\0\ and \P\1\ were cut from the same single-crystalline rod.
Typical sample dimensions were 2.0 $\times$ 0.5~mm$^2$ in the plane perpendicular to the pressure direction and 0.5~mm along the pressure direction.
The sample surfaces perpendicular to the pressure direction were polished to be parallel to each other to improve pressure homogeneity.
The side surfaces of a sample were covered with epoxy (Emerson-Cuming, Stycast 1266) to prevent the sample from breaking.
To allow the epoxy to spread freely under pressure, sufficient space was maintained between the epoxy and the pick-up coil for AC susceptibility measurement, as shown in Fig.~\ref{Coil}(a).
UAP, but not uniaxial strain, was achieved with this configuration.

UAP was applied along the [100] or [110] direction at room temperature using a piston-cylinder-type pressure cell~\cite{Kittaka2010, Taniguchi2012, Ishikawa2012, Taniguchi2013}.
No pressure medium was used.
All the inner parts were made of Cu-Be alloy, whereas the outer body was made of polybenzimidazole (hard plastic) to avoid eddy current.
The pressure was monitored using a strain gauge.

The AC susceptibility $\chi_\mathrm{AC}(T)=\chi'(T) - i\chi''(T)$ was measured 
by a mutual inductance method using a lock-in amplifier (Stanford Research Systems, SR830) and a $^3$He cryostat (Oxford Instruments, Heliox VL).
$\chi'$ was scaled from the measured pick-up-coil voltage such that $\chi'$(4~K $>$ \Tc)~=~0 and $\chi'$(0.3~K $<$ \Tc)~=~$-1$ under ambient pressure.
To improve the sensitivity by enhancing the sample filling factor, we placed a pick-up coil inside the pressure cell as shown in Fig.~\ref{Coil}.
The AC magnetic field was applied parallel to the pressure direction.
The accuracies of the field and pressure directions with respect to the crystalline axes were better than 5$^\circ$.

\section{Results}
\begin{figure}[htb]
\begin{center}
\includegraphics[width=3in]{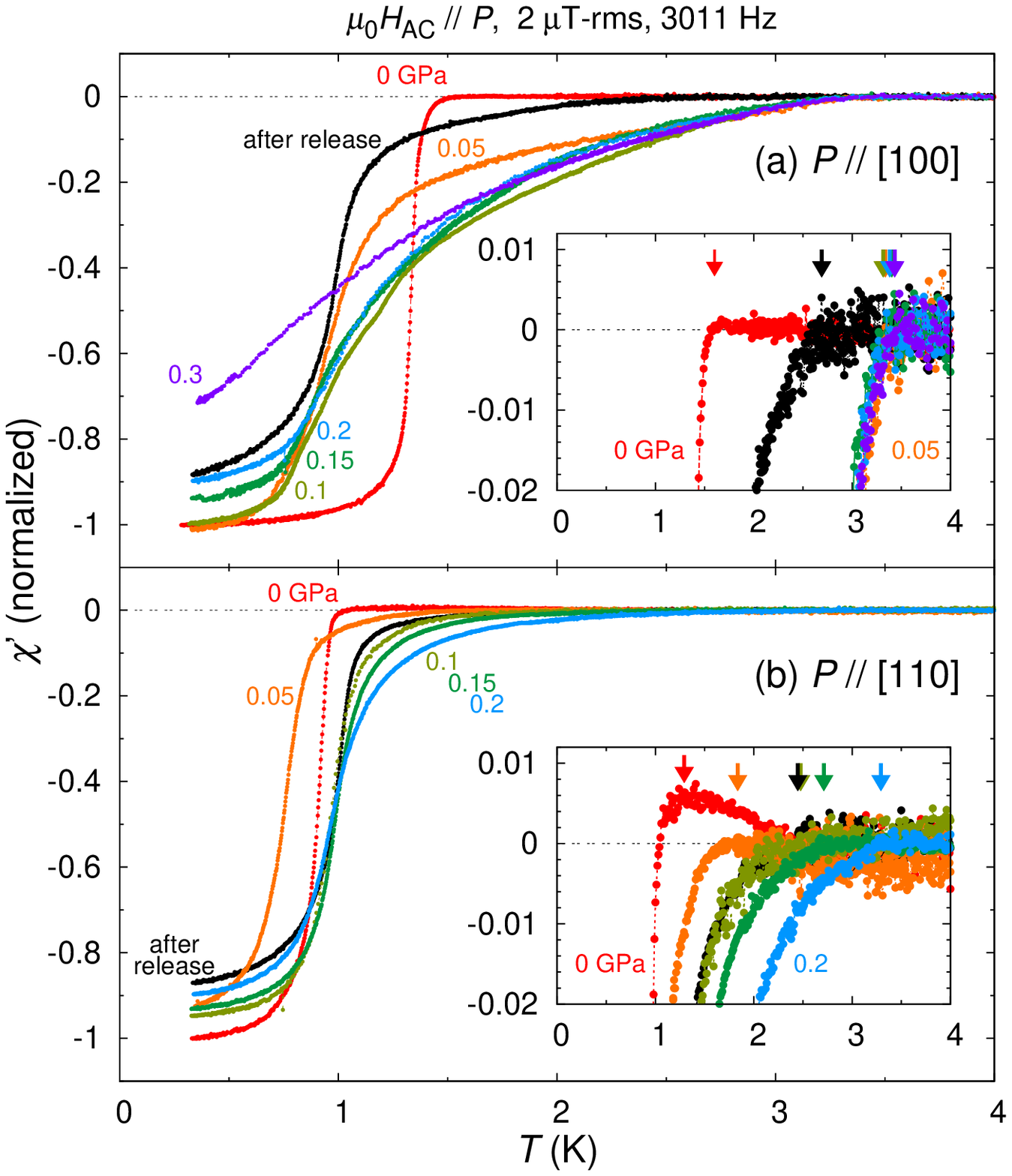}
\end{center}
\caption
{(Color online) 
Temperature dependence of the real part of the AC susceptibility of \SRO\ under (a) \P\0\ and (b) \P\1.
Black curves show the data taken about one week after releasing the pressure.
The insets are enlarged views around the SC onset.
The arrows indicate SC onset temperatures, which are defined as the temperature at which $\chi'$ starts to decrease.
$\chi'$ was scaled from the measured pick-up-coil voltage such that $\chi'$(4~K)~=~0 and $\chi'$(0.3~K)~=~$-1$ under ambient pressure.
}
\label{Re}
\end{figure}

\begin{figure}[htb]
\begin{center}
\includegraphics[width=3in]{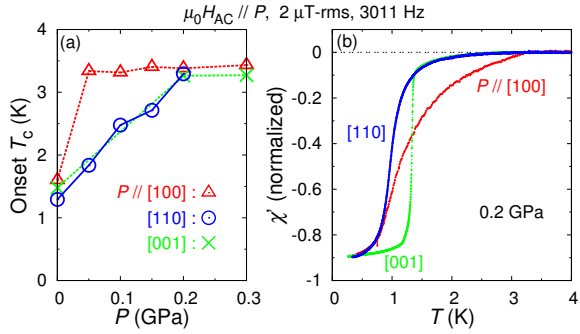}
\end{center}
\caption
{(Color online)
(a) Comparison of the pressure-magnitude dependence of the onset temperature of the superconducting transition of \SRO\ among three UAP directions, [100], [110], and [001].
The values of onset $T_{\mathrm{c}}$ are determined from AC susceptibility measurements with an AC magnetic field of 2~$\mu$T-rms and 3011~Hz parallel to the pressure direction.
(b) Comparison of the temperature dependence of the superconducting shielding fraction of \SRO\ among \P\0, \P\1, and \P\c.
}
\label{PD-dep}
\end{figure}

Figure~\ref{Re} shows the temperature dependence of the superconducting diamagnetic signal of \SRO\ under in-plane UAP.
Clearly, the diamagnetic signal above the bulk \Tc\ is nearly absent at $P=0$ but is strongly enhanced by in-plane UAP.
The onset \Tc, defined as the temperature at which the real part of the AC susceptibility, $\chi'$, starts to decrease, is enhanced to 3.3~K both by \P\0\ and \P\1.
Interestingly, we find that the UAP magnitude dependence of the onset \Tc\ is highly anisotropic, as shown in Fig.~\ref{PD-dep}(a):
the onset \Tc\ under \P\0\ is abruptly enhanced to 3.3~K at only 0.05~GPa and remains at 3.3~K for higher pressures,
whereas the onset \Tc\ under \P\1\ gradually increases and reaches 3.3~K at 0.2~GPa.

To further investigate the pressure direction dependence of the emergence of the higher-\Tc\ superconductivity, 
we plot in Fig.~\ref{PD-dep}(b) the temperature dependence of the superconducting diamagnetic signal at 0.2~GPa for \P\0\ and \P\1\ (this study), as well as for \P\c\ (from Ref.~\citenum{Kittaka2010}).
Although the higher-\Tc\ superconductivity is induced under all these pressure directions, 
the shielding fraction in the temperature range between the original bulk \Tc\ at ambient pressure and 3.3~K is markedly enhanced under \P\0.
These anisotropies in the shielding fraction and in the pressure dependence of \Tc\ indicate that \P\0\ is more effective than \P\1\ or \P\c\ in enhancing \Tc.

\begin{figure}[htb]
\begin{center}
\includegraphics[width=3in]{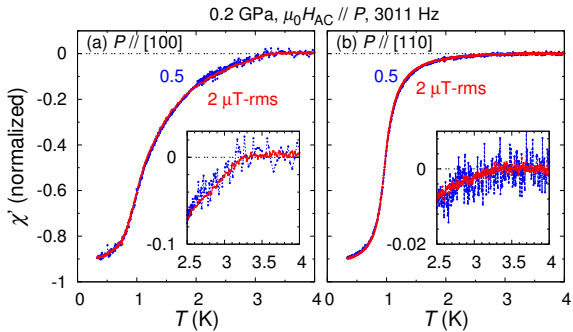}
\end{center}
\caption
{(Color online)
Superconducting diamagnetic signal in \SRO\ under (a) \P\0\ and (b) \P\1\ for $\mu_0 H_{\mathrm{AC}} = 0.5$ and 2.0~$\mu$T-rms.
The insets are enlarged views around the superconducting onset.
The two curves coincide well with each other, suggesting bulk-like 
superconductivity.
}
\label{BulkSC}
\end{figure}

To obtain information on the spatial distribution of the higher-\Tc\ superconductivity, we measured the AC-field magnitude dependence of \chiAC.
As shown in Fig.~\ref{BulkSC}, \chiAC\ for the higher-\Tc\ superconductivities induced by \P\0\ and \P\1\ is robust against changes in the amplitude of the AC magnetic field.
These results are similar to that of the higher-\Tc\ SC phase induced by out-of-plane UAP~\cite{Kittaka2010}, 
but different from that of the higher-\Tc\ SC phase in the \SRO-Ru eutectic crystal under ambient pressure~\cite{Kittaka2009JPSJ-UAP}.
In the latter, \chiAC\ is substantially decreased by strong AC-fields, probably due to the filamentary nature of superconductivity.
Thus, our present result suggests that the higher-\Tc\ superconductivity induced by in-plane UAP is not filamentary and has a bulk-like nature.

\begin{figure}[htb]
\begin{center}
\includegraphics[width=3in]{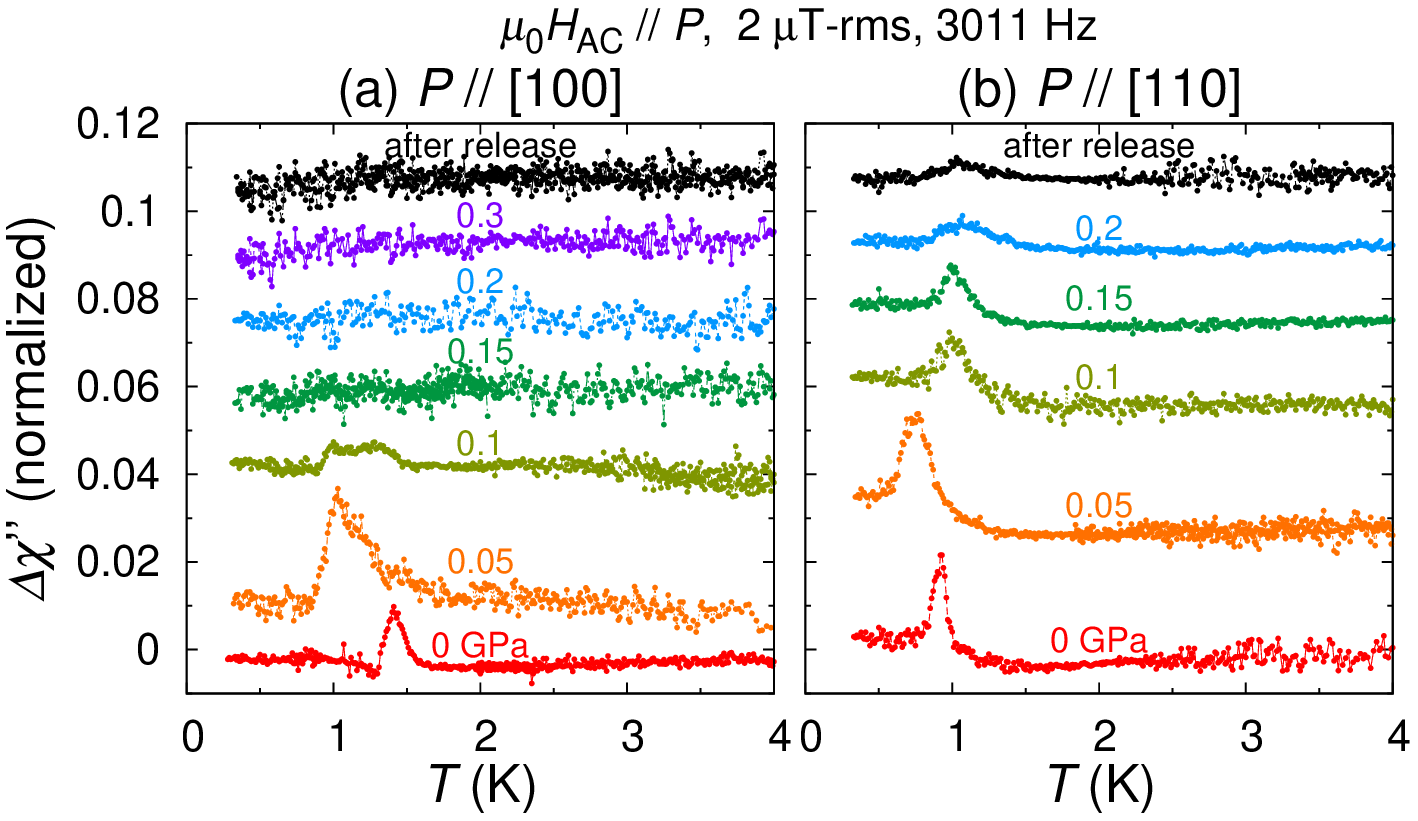}
\end{center}
\caption
{(Color online) 
Temperature dependence of the imaginary part of the AC susceptibility of \SRO\ under (a) \P\0\ and (b) \P\1.
}
\label{Im}
\end{figure}

We also measured the AC susceptibility of \SRO\ about one week after releasing the UAP from 0.3~GPa for \P\0\ and from 0.2~GPa for \P\1,
as shown in Figs.~\ref{Re} and \ref{Im}.
In the cases of both \P\0\ and \P\1, the onset \Tc\ decreases to about 2.5~K, 
and the shielding fraction is markedly suppressed in the temperature range between the original bulk \Tc\ at ambient pressure and 3.3~K.
These results indicate that the higher-\Tc\ superconductivity under in-plane UAP is induced mainly by elastic distortion,
although plastic distortion, such as dislocation~\cite{Ying2013}, should also play some role
because the onset \Tc\ after pressure release is still higher than the original value.

Figure~\ref{Im} shows the temperature dependence of the imaginary part of the AC susceptibility of \SRO\ under in-plane UAP.
As explained earlier, owing to the crystal-symmetry reduction by in-plane UAP, 
two distinct SC phases are expected to emerge depending on the temperature, and the imaginary part of the AC magnetic susceptibility might exhibit double peaks upon probing both the SC transitions.
In our experiments, only $\chi''(T)$ under \P\0\ of 0.1~GPa exhibits clear double peaks.
However, these double peaks seem to be naturally explained by the inhomogeneity of the lattice distortion over the mm-size crystal
rather than the double transitions of the distinct SC phases,
because $\chi'$ also exhibits double transitions and the peak temperatures of $\chi''$ correspond to the temperatures at which $d\chi'/dT$ becomes maximum;
in SC double transitions induced by the symmetry reduction, $\chi'(T)$ is expected to exhibit an anomaly only at the higher \Tc, \Tc$_+$.
We propose the following two possibilities as the reason why the second transition was not observed.
One possibility is that the second transition is not accompanied by any dissipation of magnetic energy.
In this case, other measurements such as specific heat are required to detect the double transitions.
The second possibility is the broadness of the actual SC transition:
a broad SC transition makes the height and width of the $\chi''$ peak smaller and larger, respectively, and results in the smearing of each peak.

\section{Discussion}
Previous studies on the \SRO-Ru eutectic, \SRO\ under out-of-plane UAP, and \SRO\ around dislocations all indicate that \SRO\ has a maximal \Tc\ of about 3~K~\cite{Kawamura2005, Kittaka2010, Ying2013}.
In this study, we have newly clarified that the higher-\Tc\ SC phase is also induced in pure \SRO\ by in-plane UAP.
As a common feature of these higher-\Tc\ SC phases, the lattice is distorted and the in-plane crystal symmetry is expected to be reduced at least in part of the sample.
Therefore, these superconductivities with \Tc\ $\sim$ 3~K emerging in the eutectic crystal, around dislocations, or under UAP are considered to have a common origin:
the higher-\Tc\ SC phase is nonchiral with a one-component order parameter, whereas the 1.5 K SC phase is chiral with a two-component order parameter~\cite{Sigrist2001}.
The optimal \Tc\ of the higher-\Tc\ SC phase is expected to be 3.3~K, because the onset \Tc\ is 3.3~K both in \SRO\ under UAP and in the \SRO-Ru eutectic.

The discrepancy between our results and the prediction based on ultrasonic experiments and the Ehrenfest relation is consistent with the scenario
that the higher-\Tc\ superconductivity induced by in-plane UAP is in a phase that is different from the phase of the 1.5 K superconductivity under ambient conditions.
The prediction, which indicates that \P\0\ suppresses \Tc, is obtained by assuming 
an elastic lattice distortion and zero-pressure limit~\cite{Okuda2002PhysicaB, Okuda2002JPSJ}.
Since the \Tc\ suppression by releasing the pressure as shown in Fig.~\ref{Re}, indicates that the main distortion is elastic,
the observed \Tc\ enhancement by \P\0\ in contrast to the prediction suggests that the situation is beyond the zero-pressure limit.
In other words, a marked change in the electronic state is expected to be accompanied by the higher-\Tc\ superconductivity:
for example, a higher-order effect becomes nonnegligible or the density of states at the Fermi level is changed.

Compared with other single-layer perovskite superconductors, the abrupt \Tc\ enhancement under the symmetry reduction of the lattice is striking.
In La$_{(2-x)}$Sr$_x$CuO$_4$, the pressure dependence of \Tc\ is consistent with the prediction based on ultrasonic experiments and the Ehrenfest relation~\cite{Nohara1995, Goko1999}.
Such a difference highlights the uniqueness of the superconductivity in \SRO: it has a two-component order parameter in the tetragonal structure, 
and orthorhombic lattice distortion leads to a new SC phase with a one-component order parameter,
whereas the superconductivity of La$_{(2-x)}$Sr$_x$CuO$_4$ remains in one phase under orthorhombic lattice distortion.

Similar to the present study, a recent study on the effect of uniaxial strain has revealed that the \Tc\ enhancement is much larger for strain along [100] than along [110]~\cite{Hicks2014}.
The symmetric change in \Tc\ between compressive and tensile [100] strains supports the scenario that the superconducting order parameter of \SRO\ has two components.
The [100] strain results shown in Ref.~\citenum{Hicks2014} at first glance appear quantitatively different from our \P\0\ results:
it was found that \Tc\ is gradually enhanced and reaches only 1.9~K at $-$0.23$\%$ strain, 
which corresponds to 0.21~GPa stress in our experiment, according to an estimation using the elastic constants in Refs.~\citenum{Okuda2002PhysicaB} and \citenum{Okuda2002JPSJ}.
We can explain this apparently different behavior by considering the difference in the experimental setups:
the pick-up coil used in our measurements surrounds the entire sample, as shown in Fig.~\ref{Coil}, and detects the whole magnetic signal originating from the sample,
whereas the pick-up coil used in Ref.~\citenum{Hicks2014} is much smaller than the sample and is located at the center of the sample,
and thus detects the susceptibility only of the central region, which is expected to be more homogeneously strained.
Thus, the superconducting signal observed in our study is dominated by regions with higher \Tc.
In fact, resistivity measurements~\cite{Hicks-R} have revealed that the sample under strain contains a part with a much higher \Tc\ than that probed by AC susceptibility.
The qualitative difference between the effects of [110] strain and pressure is also attributable to pressure inhomogeneity:
near the region close to the piston, the crystal is probably also distorted along the directions perpendicular to the pressure.
Nevertheless, the effectiveness of \P\0\ in inducing the higher-\Tc\ SC phase observed in this study is clear and consistent with the results of the uniaxial strain experiments~\cite{Hicks2014}.

Notably, the effect of \P\0\ in inducing the higher-\Tc\ SC phase is greater in pure \SRO\ (this study) than in typical \SRO-Ru eutectic crystal (Ref.~\citenum{Kittaka2009JPSJ-UAP}):
for example, the diamagnetic susceptibility at 2.5 K reaches $\sim$ 10\% at only 0.05~GPa in pure \SRO, whereas it is only 5\% at 0.45~GPa in the eutectic crystal.
These results exclude the possibility that the higher-\Tc\ SC phase in our \P\0\ experiments originates from the \SRO-Ru eutectic part in our samples which is too small to be detected under ambient conditions,
and confirm that \P\0\ can induce the higher-\Tc\ SC phase in pure \SRO.
This possibility is also excluded for the case of \P\1\ for the following reason.
If the \SRO-Ru eutectic part in the sample is assumed as the origin of the higher-\Tc\ SC phase under \P\1, the eutectic region is estimated to be as much as 2/13 of the whole sample 
by comparison of the shielding fraction at 2~K under \P\1\ of 0.2~GPa between the eutectic sample described in Ref.~\citenum{Kittaka2009JPSJ-UAP} and our \SRO\ sample.
However, this estimated value is less likely, 
because no Ru inclusion is observed on the sample surface under an optical microscope and no higher-\Tc\ superconductivity emerges under ambient pressure.

Next, we discuss the origin of the difference between the effects of \P\0\ and \P\1\ on the evolution of the higher-\Tc\ SC phase.
For \P\0, half of the in-plane Ru-O bonds are parallel to the pressure direction and the other half are perpendicular to it,
whereas for \P\1, all the in-plane Ru-O bonds are at 45$^\circ$ to the pressure direction.
The direct compression by \P\0\ of the Ru-O bonds along the pressure direction will result in the substantial shortening of these in-plane Ru-O bond lengths and \Tc\ enhancement with smaller pressure.
As another aspect of the effect of \P\0, it has been pointed out from the band calculation that the $\gamma$ Fermi surface, which probably plays an active role in the superconductivity of \SRO, 
approaches to the van Hove singularity at the M point~\cite{Hicks2014}.
This tendency should also be important for the emergence of the higher-\Tc\ SC phase under \P\0\ of only 0.05~GPa.
As a future project, the measurement of de Haas-van Alphen oscillations is a promising technique for following the evolution of the Fermi surfaces of \SRO\ under in-plane UAP.

The anisotropic pressure dependence of \Tc, $\partial T_\mathrm{c} / \partial P$, under \P\0\ and \P\1, is related to coefficients in the Ginzburg-Landau free energy~\cite{Walker2002}.
The Gibbs free energy $F_{\mathrm{GL}}$ and the coupling energy $F_{\mathrm{couple}}$ between the crystal lattice deformation and the SC order parameter
are described as follows:
\begin{equation} 
\begin{aligned}
F_{\mathrm{GL}} = & \alpha ' (T - T_{\mathrm{c}0}) (|\Psi_x|^2 + |\Psi_y|^2) + \frac{b_1}{4} (|\Psi_x|^2 + |\Psi_y|^2)^2 \\
& + b_2 |\Psi_x|^2 |\Psi_y|^2 + \frac{b_3}{2} (\Psi_x^2 \Psi_y^{*2} + \Psi_y^2 \Psi_x^{*2}) \; ,
\end{aligned}
\end{equation}
\begin{equation}
F_{\mathrm{couple}} = \sigma_i d_{ij} E_j \; .
\end{equation}
In the above description, a two-component SC order parameter ($\Psi_x$, $\Psi_y$) is assumed and $T_{\mathrm{c}0}$ is the transition temperature under ambient pressure.
$\alpha '$ is a coefficient with a positive sign to express the temperature-linear term of $F_{\mathrm{GL}}$.
$\sigma_i$ are the stress tensor components ($i$ = 1, ... 6 using the Voigt notation).
The matrix $d_{ij}$ has the same symmetry properties as the elastic compliance matrix $s_{ij}$ appropriate for tetragonal symmetry.
The nonzero components of $E_j$ are $E_1 = |\Psi_x|^2$, $E_2 = |\Psi_y|^2$, and $E_6 = \Psi_x^* \Psi_y + \Psi_x \Psi_y^*$.
By minimizing the sum of these energies, $F_{\mathrm{GL}} + F_{\mathrm{couple}}$, 
the transition temperature of the higher-\Tc\ SC phase with a one-component order parameter \Tc$_+$ is obtained as a function of the magnitude $\sigma$ of compression stress ($\sigma>$ 0):
\begin{equation}
T_{\mathrm{c}+} = T_\mathrm{c0} - \frac{\sigma}{\alpha '}d_{11} \;\;\; (\mathrm{under} \; P_{||[100]}) \; ,
\end{equation}
\begin{equation}
T_{\mathrm{c}+} = T_\mathrm{c0} - \frac{\sigma}{2 \alpha '}(d_{11} + d_{12} +d_{66}) \;\;\; (\mathrm{under} \; P_{||[110]}) \; .
\end{equation}
The large \Tc\ enhancement under \P\0\ indicates that $d_{11} < 0$,
whereas the small \Tc\ suppression under \P\1, which is considered to be intrinsic, indicates that $(d_{11} + d_{12} + d_{66}) \geq 0$.
From these relations, we can obtain
\begin{equation}
- d_{11} \leq d_{12} +d_{66} \; .
\end{equation}
For further quantitative discussion, the effects of UAP for different pressure directions should be measured using samples with the same \Tc.
Note that the original values of $T_\mathrm{c}$ were different among the samples used for the \P\0, \P\1, and \P\c\ experiments.

Lastly, we discuss the origin of the strange upturn behavior between 2.4 and 1.3~K observed in $\chi'$ under \P\1\ of 0~GPa, which is shown in the inset of Fig.~\ref{Re}(b).
It is unlikely to be due to Langevin-type paramagnetic impurities for the following reasons.
The amount of assumed paramagnetic impurities from fitting the susceptibility data is about $10^5$~ppm.
This value is unrealistically large for our samples since the upper bound of the amount of impurities is estimated to be about 700~ppm from its bulk \Tc,
which is 0.6~K lower than the optimal \Tc\ of 1.5~K~\cite{Kikugawa2002}.
We suggest the possibility that a magnetic impurity phase macroscopically exists in this sample.

\section{Summary}
We measured the AC susceptibility of \SRO\ under in-plane UAP
and have clarified that the onset temperature of the superconducting transition is enhanced to 3.3~K both by \P\0\ and \P\1.
The maximal \Tc\ of 3.3~K is obtained at only 0.05~GPa under \P\0, whereas 0.2~GPa is required to induce the same \Tc\ under \P\1.
The fact that \P\0\ is more effective in inducing a higher-\Tc\ SC phase is considered to be closely related to
the direct shortening of the Ru-O bond length and the approach of the $\gamma$ Fermi surface to the van Hove singularity under \P\0.

\section*{}
\begin{acknowledgments}
We thank C. W. Hicks, A. P. Mackenzie, T. Yamazaki, and H. Yaguchi for fruitful discussion.
This work was supported
by the Ministry of Education, Culture, Sports, Science and Technology (MEXT) KAKENHI (Nos. 25610095 and 22103002)
and by a Grant-in-Aid for the Global COE program ``The Next Generation of Physics, Spun from Universality and Emergence'' from MEXT, Japan.
H. T. acknowledges the Japan Society for the Promotion of Science for a fellowship (No. 12J01430) and the University of Cambridge for hospitality.
S. K. G. thanks Trinity College for a fellowship.
\end{acknowledgments}

\bibliography{string,Ca2RuO4,Sr2RuO4,others}

\begin{thebibliography}{10}

\bibitem{Maeno1994}
Y.~Maeno, H.~Hashimoto, K.~Yoshida, S.~Nishizaki, T.~Fujita, J.~G. Bednorz, and
  F.~Lichtenberg, Nature {\bfseries 372}, 532 (1994).

\bibitem{Rice1995}
T.~M. Rice and M.~Sigrist, J. Phys.: Condens. Matter {\bfseries 7}, L643
  (1995).

\bibitem{Ishida1998}
K.~Ishida, H.~Mukuda, Y.~Kitaoka, K.~Asayama, Z.~Q. Mao, Y.~Mori, and Y.~Maeno,
  Nature {\bfseries 396}, 658 (1998).

\bibitem{Mackenzie1998PRL}
A.~P. Mackenzie, R.~K.~W. Haselwimmer, A.~W. Tyler, G.~G. Lonzarich, Y.~Mori,
  S.~Nishizaki, and Y.~Maeno, Phys. Rev. Lett. {\bfseries 80}, 161 (1998).

\bibitem{Duffy2000}
J.~A. Duffy, S.~M. Hayden, Y.~Maeno, Z.~Mao, J.~Kulda, and G.~J. McIntyre,
  Phys. Rev. Lett. {\bfseries 85}, 5412 (2000).

\bibitem{Ishida2001}
K.~Ishida, H.~Mukuda, Y.~Kitaoka, Z.~Q. Mao, H.~Fukazawa, and Y.~Maeno, Phys.
  Rev. B {\bfseries 63}, 060507 (2001).

\bibitem{Maeno2012}
Y.~Maeno, S.~Kittaka, T.~Nomura, S.~Yonezawa, and K.~Ishida, J. Phys. Soc. Jpn.
  {\bfseries 81}, 011009 (2012).

\bibitem{Luke1998}
G.~M. Luke, Y.~Fudamoto, K.~M. Kojima, M.~I. Larkin, J.~Merrin, B.~Nachumi,
  Y.~J. Uemura, Y.~Maeno, Z.~Q. Mao, Y.~Mori, H.~Nakamura, and M.~Sigrist,
  Nature {\bfseries 394}, 558 (1998).

\bibitem{Xia2006}
J.~Xia, Y.~Maeno, P.~T. Beyersdorf, M.~M. Fejer, and A.~Kapitulnik, Phys. Rev.
  Lett. {\bfseries 97}, 167002 (2006).

\bibitem{Agterberg1998}
D.~F. Agterberg, Phys. Rev. Lett. {\bfseries 80}, 5184 (1998).

\bibitem{Sigrist2002}
M.~Sigrist, Prog. Theor. Phys. {\bfseries 107}, 917 (2002).

\bibitem{Nishizaki2000}
S.~NishiZaki, Y.~Maeno, and Z.~Mao, J. Phys. Soc. Jpn. {\bfseries 69}, 572
  (2000).

\bibitem{Mao2000PRL}
Z.~Q. Mao, Y.~Maeno, S.~NishiZaki, T.~Akima, and T.~Ishiguro, Phys. Rev. Lett.
  {\bfseries 84}, 991 (2000).

\bibitem{Tanatar2001PRB}
M.~A. Tanatar, S.~Nagai, Z.~Q. Mao, Y.~Maeno, and T.~Ishiguro, Phys. Rev. B
  {\bfseries 63}, 064505 (2001).

\bibitem{Deguchi2002}
K.~Deguchi, M.~A. Tanatar, Z.~Mao, T.~Ishiguro, and Y.~Maeno, J. Phys. Soc. Jpn
  {\bfseries 71}, 2839 (2002).

\bibitem{Yaguchi2002}
H.~Yaguchi, T.~Akima, Z.~Mao, Y.~Maeno, and T.~Ishiguro, Phys. Rev. B
  {\bfseries 66}, 214514 (2002).

\bibitem{Tenya2006}
K.~Tenya, S.~Yasuda, M.~Yokoyama, H.~Amitsuka, K.~Deguchi, and Y.~Maeno, J.
  Phys. Soc. Jpn. {\bfseries 75}, 023702 (2006).

\bibitem{Yonezawa2013}
S.~Yonezawa, T.~Kajikawa, and Y.~Maeno, Phys. Rev. Lett. {\bfseries 110},
  077003 (2013).

\bibitem{Yonezawa2014}
S.~Yonezawa, T.~Kajikawa, and Y.~Maeno, J. Phys. Soc. Jpn. {\bfseries 83},
  083706 (2014).

\bibitem{Shirakawa1997}
N.~Shirakawa, K.~Murata, S.~Nishizaki, Y.~Maeno, and T.~Fujita, Phys. Rev. B
  {\bfseries 56}, 7890 (1997).

\bibitem{Forsythe2002}
D.~Forsythe, S.~R. Julian, C.~Bergemann, E.~Pugh, M.~J. Steiner, P.~L. Alireza,
  G.~J. McMullan, F.~Nakamura, R.~K.~W. Haselwimmer, I.~R. Walker, S.~S.
  Saxena, G.~G. Lonzarich, A.~P. Mackenzie, Z.~Q. Mao, and Y.~Maeno, Phys. Rev.
  Lett. {\bfseries 89}, 166402 (2002).

\bibitem{Okuda2002PhysicaB}
N.~Okuda, T.~Suzuki, Z.~Mao, Y.~Maeno, and T.~Fujita, Physica B {\bfseries
  312}, 800 (2002).

\bibitem{Okuda2002JPSJ}
N.~Okuda, T.~Suzuki, Z.~Mao, Y.~Maeno, and T.~Fujita, J. Phys. Soc. Jpn.
  {\bfseries 71}, 1134 (2002).

\bibitem{Maeno1998}
Y.~Maeno, T.~Ando, Y.~Mori, E.~Ohmichi, S.~Ikeda, S.~NishiZaki, and
  S.~Nakatsuji, Phys. Rev. Lett. {\bfseries 81}, 3765 (1998).

\bibitem{Ando1999}
T.~Ando, T.~Akima, Y.~Mori, and Y.~Maeno, J. Phys. Soc. Jpn. {\bfseries 68},
  1651 (1999).

\bibitem{Kittaka2009JPSJ-SCregion}
S.~Kittaka, T.~Nakamura, H.~Yaguchi, S.~Yonezawa, and Y.~Maeno, J. Phys. Soc.
  Jpn. {\bfseries 78}, 064703 (2009).

\bibitem{Kittaka2010}
S.~Kittaka, H.~Taniguchi, S.~Yonezawa, H.~Yaguchi, and Y.~Maeno, Phys. Rev. B
  {\bfseries 81}, 180510(R) (2010).

\bibitem{Ikeda2001}
S.~Ikeda, N.~Shirakawa, S.~Koiwai, A.~Uchida, M.~Kosaka, and Y.~Uwatoko,
  Physica C {\bfseries 364}, 376 (2001).

\bibitem{Yaguchi-2014JPS}
Y.~Takanuki, T.~Yamazaki, and H.~Yaguchi, presented at JPS 69th Annual Meeting,
  2014.

\bibitem{Hicks2014}
C.~W. Hicks, D.~O. Brodsky, E.~A. Yelland, A.~S. Gibbs, J.~A.~N. Bruin, M.~E.
  Barber, S.~D. Edkin, K.~Nishimura, S.~Yonezawa, Y.~Maeno, and A.~P.
  Mackenzie, Science {\bfseries 344}, 283 (2014).

\bibitem{Mao2000MRB}
Z.~Q. Mao, Y.~Maeno, and H.~Fukazawa, Mater. Res. Bull. {\bfseries 35}, 1813
  (2000).

\bibitem{Taniguchi2012}
H.~Taniguchi, S.~Kittaka, S.~Yonezawa, H.~Yaguchi, and Y.~Maeno, J. Phys.:
  Conf. Ser. {\bfseries 391}, 012108 (2012).

\bibitem{Ishikawa2012}
R.~Ishikawa, H.~Taniguchi, S.~K. Goh, S.~Yonezawa, F.~Nakamura, and Y.~Maeno,
  J. Phys.: Conf. Ser. {\bfseries 400}, 022036 (2012).

\bibitem{Taniguchi2013}
H.~Taniguchi, K.~Nishimura, R.~Ishikawa, S.~Yonezawa, S.~K. Goh, F.~Nakamura,
  and Y.~Maeno, Phys. Rev. B {\bfseries 88}, 205111 (2013).

\bibitem{Kittaka2009JPSJ-UAP}
S.~Kittaka, H.~Yaguchi, and Y.~Maeno, J. Phys. Soc. Jpn. {\bfseries 78}, 103705
  (2009).

\bibitem{Ying2013}
Y.~A. Ying, N.~E. Staley, Y.~Xin, K.~Sun, X.~Cai, D.~Fobes, T.~Liu, Z.~Mao, and
  Y.~Liu, Nat. Commun. {\bfseries 4}, 2596 (2013).

\bibitem{Kawamura2005}
M.~Kawamura, H.~Yaguchi, N.~Kikugawa, Y.~Maeno, and H.~Takayanagi, J. Phys.
  Soc. Jpn. {\bfseries 74}, 531 (2005).

\bibitem{Sigrist2001}
M.~Sigrist and H.~Monien, J. Phys. Soc. Jpn. {\bfseries 70}, 2409 (2001).

\bibitem{Nohara1995}
M.~Nohara, T.~Suzuki, Y.~Maeno, T.~Fujita, I.~Tanaka, and H.~Kojima, Phys. Rev.
  B {\bfseries 52}, 570 (1995).

\bibitem{Goko1999}
T.~Goko, F.~Nakamura, and T.~Fujita, J. Phys. Soc. Jpn. {\bfseries 68}, 3074
  (1999).

\bibitem{Hicks-R}
{C}. Hicks, private communication.

\bibitem{Walker2002}
M.~B. Walker and P.~Contreras, Phys. Rev. B {\bfseries 66}, 214508 (2002).

\bibitem{Kikugawa2002}
N.~Kikugawa and Y.~Maeno, Phys. Rev. Lett. {\bfseries 89}, 117001 (2002).

\end{thebibliography}

\end{document}